\documentstyle[preprint,prl,aps,psfig,amssymb]{revtex}
\begin{document}
% \twocolumn[\hsize\textwidth\columnwidth\hsize\csname @twocolumnfalse\endcsname
\draft
\begin{title}
 {  Weak-localization and Rectification Current 
in Non-diffusive Quantum Wires }
\end{title}
\author{ Kang-Hun Ahn}
\address{ Physics Research Division and Center for Theoretical Physics, \\ School of Physics, Seoul National
University, Seoul 151-747, Korea }

\date{\today}
\maketitle

\widetext

\begin{abstract}
We show that electron transport in disordered quantum wires can be
described by a modified Cooperon equation, which coincides in form
with the Dirac equation for the massive fermions in a 1+1 dimensional
system. In this new formalism, we calculate the DC electric current
induced by electromagnetic fields in quasi-one-dimensional rings. This
current changes sign, from diamagnetic to paramagnetic, depending on
the amplitude and frequency of the time-dependent external
electromagnetic field.
\end{abstract}
%\vspace{0.2cm}
%03.65.-w Quantum mechanics
%72.15.-v Electronic conduction in metals and alloys
%73.23.-b Electronic transport in mesoscopic systems

%]
 \narrowtext

Impurity scattering and quantum coherence of the electron 
wavefunction are the two key concepts in the transport phenomena of mesoscopic
conductors. Conduction electrons are weakly localized by the coherent 
back-scattering due to impurities, resulting in the effect commonly known as 
``weak localization". The conventional theory of weak 
localization\cite{altshuler} assumes deeply diffusive systems--
i.e., the electron mean free path $l$ is much shorter than the system size.
Recently,  experimental study of 
transport phenomena in non-diffusive systems
% ($l\stackrel{>}{\sim} L$)have also become important, primarily
have also become important, primarily
due to the recent progress in fabrication of clean nanostructures.
In this paper, we  present a formalism of the weak localization phenomena,
valid also in non-diffusive regimes.
%The electron transport in quantum wires will be shown to be mapped into
%a Dirac theory with imaginary time for a massive particle in 1+1 dimension.
In particular, we consider electromagnetic(EM)-field-induced  current in mesoscopic
rings, which is currently an important issue  concerning the sign of the 
measured persistent current\cite{persistent,jariwala,deblock}.

% This current displays interesting features due to the combination of nonlinear
%and ballistic effects. 

Nonlinear properties of field-induced current in mesoscopic rings have 
been studied in great detail 
for the case of deeply diffusive regime\cite{kravtsov1,aronov}.
This problem has recently regained attention due to its
relevance to the problem of anomalously large persistent
current\cite{persistent,jariwala,deblock}
and low-temperature saturation of decoherence time\cite{mohanty2,mohanty1,kravtsov2}.
 We investigate the same physical model without using diffusion
approximation which is valid only for $l\ll L$.
The particular system considered in this paper is a quantum wire in ring
geometry
with finite width much larger than the Fermi wavelength but smaller than
the phase coherence length.
We show that rectified DC currents in mesoscopic rings induced by
high-frequency magnetic fields have oscillating sign depending on
the frequency. This result sheds some light on the recent puzzle on the
measured sign of the induced DC current in mesoscopic quantum rings
\cite{jariwala,deblock,kravtsov2}.

We start with the conventional weak localization theory.
Central to quantum transport in
 disordered conductors is the concept of so-called "Cooperon", the  
particle-particle diffusion propagator \cite{altshuler,montambaux}.
 The Cooperon is a two-particle Green function averaged over disorder
configurations.
In the presence of an electromagnetic field ${\bold A}$, the Cooperon is
the retarded classical propagator of a modified diffusion equation:
\begin{eqnarray}
\left[\frac{\partial}{\partial t} - D \left(\nabla_{\bold r}-\frac{2ie}{\hbar c}{\bold A}\right)^{2}+\frac{1}{\tau_{\phi}}\right]
 C({\bold r},t;{\bold r^{\prime}},t^{\prime})=\delta ({\bold
r}- {\bold r^{\prime}}) \delta(t-t^{\prime}), \label{coop}
\end{eqnarray}
where $D=v_{F}l/d$ is the diffusion coefficient for the d-dimensional system, 
$v_{F}$ is the Fermi velocity, and $\tau_{\phi}$ is the phase coherence time.
This expression in Eq.(\ref{coop}) has proven to be useful in many
cases, since one can easily consider geometrical effects through the boundary
condition of the equation. However, it is worthwhile to 
note that Eq.(\ref{coop}) is only valid in the deeply diffusive regime.
Even its original derivation hinged on
the realization that the Fourier transformed Cooperon $C({\bold Q},\omega)$
could be approximated as\cite{altshuler,bergmann}
$C({\bold Q},\omega) \approx 1/(-i\omega +DQ^{2})$ when
\begin{eqnarray}
Q l << 1 {\rm ~and ~} \omega \tau << 1,
\label{diff-limit}
\end{eqnarray}
where $\tau = l/v_{F}$ is the elastic mean free time.

When $\omega$ is not much small compared to $1/\tau$,
 one relies on the semiclassical Boltzmann theory\cite{chakravarty} instead of Eq.(\ref{coop}) for the Cooperon; the semiclassical Boltzmann theory is free from the above constraining 
approximation in Eq.(\ref{diff-limit}).
In this theory\cite{chakravarty}, the electron motion is characterized by
a function
 $F({\bold r},{\bold v},t;{\bold r^{\prime}},{\bold v^{\prime}},t^{\prime})$
 which is a conditional probability density
for the particle initially at position ${\bold r}^{\prime}$ and time $t^{\prime}$ with velocity ${\bold v^{\prime}}$ to
be found at the position ${\bold r}$ and time $t$ with the velocity ${\bold v}$.
The intrinsic velocity of the particle is fixed to be Fermi velocity,
 $|{\bold v}|=|{\bold v^{\prime}}|=v_{F}$. 
 The conditional probability density $F$ is the propagator of the 
 distribution function 
$f({\bold r},{\bold v},t)$ which satisfy the Boltzmann equation 
\begin{eqnarray}
 \label{semicl}
 \left[
\frac{\partial}{\partial t}+
{\bold v}\cdot
\left(
\nabla_{\bold r}- \frac{2ie}{\hbar c}  {\bold A}
 \right)
\right]f
&=& -\frac{f-f_{0}}{\tau} -\frac{f}{\tau_{\phi}} \\
f_{0}({\bold r},t)&=&\frac{ 1}{\mathcal N}\sum_{\bold v}f({\bold r},{\bold v},t)
\end{eqnarray}
 where ${\mathcal N}$ is the number of available values of ${\bold v}$. 

 From this point onwards, we will consider mesoscopic quantum wire 
with finite width $W$ in two dimension.
The boundary condition 
is that $f({\bold r},{\bold v},t)$ is zero if $v_{y} \neq 0 $ 
at the boundary of the wire 
$ {\bold r}=x \hat{x} \pm W/2 \hat{y}$.   
This condition is due to the fact that the electron number is conserved in
the electron scattering with the boundary.
We impose another condition on the  
width of the wire $W$ is small enough that $W/\tau_{\phi} << v_{F}$.
In this case, the main contribution to the electron propagator $F$
is essentially zero mode in the transverse direction, i. e., there is
no $y$-dependence in $F$.
To be consistent with the boundary condition, $v_{y}=0$ at the boundaries,
we have only two values of ${\bold v}$ in the longitudinal direction,
 either $v_{F}\hat{x}$ or $-v_{F}\hat{x}$. 
The equation of the propagator for the Boltzmann equation (\ref{semicl})
can be rewritten as a differential equation in a 2 by 2 matrix form:
\begin{eqnarray}
\label{main}
 \big[\frac{\partial}{\partial t}&+&v_{F}\sigma_{z}
\left(\frac{\partial}{\partial x} -\frac{2ie}{\hbar c}A \right)
+\frac{1}{2\tau}\big(1-\sigma_{x}\big)+\frac{1}{\tau_{\phi}}\big]
    {\mathcal \bold F}(x,t;x^{\prime},t^{\prime})
    \\ \nonumber
    &=&\delta(x-x^{\prime})\delta(t-t^{\prime}),
\end{eqnarray}
where
\begin{eqnarray}
\nonumber
&&{\bold F}(x,t;x^{\prime},t^{\prime}) \\
&=&\left(\begin{array}{rr} F(x,v_{F},t;x^{\prime},v_{F},t^{\prime})&F(x,v_{F},t;x^{\prime},-v_{F},t^{\prime})
\\F(x,-v_{F},t;x^{\prime},v_{F},t^{\prime})&F(x,-v_{F},t;x^{\prime},-v_{F},t^{\prime})\end{array}\right).
\label{F_def}
\end{eqnarray}

As it is clear in Eq.(\ref{F_def}), each component of the above matrix formalism denotes the relevant
chirality of the moving particle.
Since the Cooperon $C(x,t;x^{\prime},t^{\prime})$ is also a probability density
for the particle initially at $(x^{\prime}$,$t^{\prime})$  
to be found at $x$ after time
$t-t^{\prime}$\cite{montambaux,chakravarty},
we get the Cooperon from ${\bold F}$
 by summing all
final chiral states and averaging over initial chiral states; $C(x,t;x^{\prime},t^{\prime})
=\frac{1}{2} \sum_{ij}{\bold F}_{ij}(x,t;x^{\prime},t^{\prime})$.
In a matrix form, we get
\begin{eqnarray}
C(x,t;x^{\prime},t^{\prime})
 ={\rm Tr}\left[ \frac{1+\sigma_{x}}{2} { \bold
 F}(x,t;x^{\prime},t^{\prime})\right].
\label{projection}
\end{eqnarray}

Note that the main equation in Eq.(\ref{main}) coincides
in form with  a Dirac equation for a massive particles in 1+1 dimension.
Interestingly, it is already well known that
the conventional Cooperon equation (\ref{coop}) coincides in form with the
Schr\"odinger equation with imaginary time
$t \longleftrightarrow -it$
for a particle with mass
$m^{\prime} \longleftrightarrow \hbar/2D$.
 To be more explicit, let us consider the following relativistic propagator $G^{rel}$ for the Dirac equation
for the particle with mass $m^{\prime}$ in 1+1 dimension:
\begin{eqnarray}
\nonumber
&&\big[i\gamma^{0}\left( \frac{\partial}{\partial t} +i\frac{m^{\prime}c^{2}}{\hbar} \right)
-ic\gamma^{1}
\left(\frac{\partial}{\partial x}-\frac{2ie}{\hbar c}A \right)
+\frac{m^{\prime}c^{2}}{\hbar}  \big]G^{rel}
\\
&~~~~~&=i\gamma^{0}\delta(x-x^{\prime})\delta(t-t^{\prime}),
\label{relativistic}
\end{eqnarray}
where $\{\gamma^{\mu},\gamma^{\nu}\}=2g^{\mu \nu}$ and $g^{\mu \nu}={\rm diag}(1,-1)$ for $\mu,\nu=0,1$.
The term $i\frac{m^{\prime}c^{2}}{\hbar} $
 within the first parenthesis is included to compensate
the time-evolution factor $\exp(-i \frac{m^{\prime}c^{2}}{\hbar}t)$
due to the rest-mass energy which is not included in the Schr\"odinger
equation.
Let us perform the following transformations, while keeping
 the electron coupling constant $\frac{2ie}{\hbar c}$ to the EM
 field,
\begin{eqnarray}
t\rightarrow -it,~ c\rightarrow i v_{F},~ m^{\prime}\rightarrow \hbar/2D,
\label{replace}
\end{eqnarray}
then we get
Eq.(\ref{main}) using Pauli spin matrices
$\gamma^{0}=\sigma_{x}$ and $\gamma^{1}=i\sigma_{y}$.

The physical origin of the transformation $c\rightarrow iv_{F}$ in Eq.(\ref{replace}) 
is rendered clear by noting 
that the mean speed of a particle in a disordered conductor
is limited by the Fermi velocity while the speed of a particle in the relativistic
theory can not exceed the speed of light.
Instead of $\sigma_{x}$ chosen as $\gamma^{0}$ in the above, 
a general form of Eq.(\ref{projection}) can be used using $\gamma$
matrices;
$C(x,t;x^{\prime},t^{\prime})
 ={\rm Tr}\left[ \left(\frac{1+\gamma^{0}}{2}\right) {\bold F}(x,t;x^{\prime},t^{\prime})\right]$.
The appearance of
  $(1+\gamma^{0})/2$ in front of ${\bold F}$ seems to be
natural, because $(1+\gamma^{0})/2$ in relativistic mechanics
  is a projection operator, which projects out negative energy states 
in the rest reference frame\cite{ryder}.

Pursuing further similarity between the present theory and
relativistic theory, let us note
how the chiral symmetry is broken in each theory.
 In relativistic quantum mechanics in 1+1 dimension, the particle's mass breaks the chiral symmetry.
(In other words,  since the massive particle moves slower than $c$, there exist reference frames where a right-moving particle can be seen as a left-moving particle.)
In mesoscopic quantum wires, the chirality is broken due to the presence of
the electron-impurity scattering.
These two different mechanisms, which break the chiral symmetry
in each theory, are  connected to each other as clearly manifested by the correspondence
shown in Eq.(\ref{replace});
$ m^{\prime} c^{2}  \longrightarrow  -\hbar v_{F}^{2} /2D =-\hbar/2\tau $.

In the absence of external fields, ${\bold F}$ is a translationally invariant
quantity, which allows us to solve Eq.(\ref{main})  using Fourier
transform:
${\bold F}(x,t;x^{\prime},t^{\prime})=(1/2\pi)^{2}\int dQ \int d\omega
{\bold F}(Q,\omega)e^{iQ (x-x^{\prime})-i\omega (t-t^{\prime})}$.
For $\tau_{\phi} >> \tau$, ${\bold F}(Q,\omega)$ is given by
\begin{eqnarray}
\nonumber
 {\bold F}(Q,\omega)&=&\frac{1}{-i\omega+DQ^{2}-\omega^{2}\tau}
\\&&\times \left(
\begin{array}{rr} 1/2-i\omega\tau- i l Q & 1/2 ~~~~~~~~~\\ 1/2~~~~~~~~~& 1/2-i\omega\tau+ i l Q \end{array}
\right).
\end{eqnarray}
The Cooperon in momentum space is written as
\begin{eqnarray}
C(Q,\omega)={\rm Tr}\frac{1+\sigma_{x}}{2}{\bold F}(Q,\omega)
=\frac{1-i\omega\tau }{-i\omega+DQ^{2}-\omega^{2}\tau}.
\end{eqnarray}
This result coincides with the Cooperon obtained as the total sum of the Dyson series
in the Green function approach
without the approximations in Eq.(\ref{diff-limit})\cite{ater}.

In the presence of time-dependent EM field, ${\bold A}=A(t)\hat{x}$,
the ``Cooperon matrix" ${\bold F}$
explicitly depends on time $t$
    (${\bold F}_{t}={ \bold F}_{t}(x,\eta;x^{\prime},\eta^{\prime})$),
which is obtained by
solving the following equation with time-dependent field
$A_{t}(\eta)=A(t-\eta/2)+A(t+\eta/2)$\cite{altshuler};
\begin{eqnarray}
\nonumber
&& \big[\frac{\partial}{\partial \eta}+v_{F}\sigma_{z}\left(\frac{\partial}{\partial
x}-\frac{ie}{\hbar c}A_{t}(\eta)\right)
+\frac{1-\sigma_{x}}{2\tau}+\frac{1}{\tau_{\phi}^{*}}\big]
    { \bold F}_{t} \\
    &=&\delta(x-x^{\prime})\delta(\eta-\eta^{\prime}),
\label{main2}
\end{eqnarray}

Here, we include the phenomenological
dephasing rate $1/\tau_{\phi}^{*}$, which has its origin from sources other than 
the external EM field.
The weak localization current $I_{WL}(t)$\cite{altshuler} (quantum correction to classical ohmic current)
is given by
\begin{eqnarray}
\nonumber
 \left< I_{WL}(t) \right>&=&
\nonumber
 \frac{C_{\beta}e^{2}D}{ \hbar}
\int_{0}^{\infty}d\eta
 {\rm Tr}
\left[\frac{1+\sigma_{x}}{2}{ {\bold F}}_{t-\eta/2}(x,\eta;x,-\eta)
\right]\\ &\times& E(t-\eta),
\label{main3}
\end{eqnarray}
where $\left<\cdots\right>$ represents disorder average and $E(t)=-\frac{1}{c}\frac{\partial A(t)}{ \partial t} $ is the applied electric field.
$C_{\beta}$ is dictated by the Dyson symmetry class ;
$C_{\beta}=-4/\pi$ ($2/\pi$) when the spin-orbit scattering is negligible (important)  with the characteristic length
$L_{so} >> L$ ($L_{so} << L$)\cite{altshuler,kravtsov2}.

Now, let us apply Eq.(\ref{main2}) and Eq.(\ref{main3}) to
calculate
electric currents induced by the EM field in mesoscopic rings.
While the usual equilibrium persistent current is induced by
a static magnetic flux $\phi=\bar{A}L$ only,
the rectified direct current is a dynamical phenomenon
originating from the time-dependent conductivity of the ring\cite{kravtsov1,kravtsov2},
Suppose the EM field, given by $A(t)=\bar{A}+a(t)$, is applied to the quantum ring
with perimeter $L$, where
$a(t)=\frac{1}{2}(a_{\omega}e^{-i\omega t}+c.c.)$ and $\bar{A}$ is time-independent.
An electric field
 ${\mathcal E}(t)=\frac{1}{2}({\mathcal E}_{\omega}e^{-i\omega t} +c.c)$ is induced along the ring, where
${\mathcal E}_{\omega}=i\omega a_{\omega} /c$.
The DC component of electric current $I_{0}=\overline{\left<I_{WL}(t)\right>}$ is of interest,
and it is obtained by averaging the disorder-averaged current $\left<I_{WL}(t)\right>$  over time $t$.

Let us first investigate the case of  weakly time-dependent field so
that the associated magnetic flux $\phi_{\omega}$ is much smaller
than the unit flux quantum $\phi_{0}=h/|e|c$;
\begin{eqnarray}
\phi_{\omega}
=|{\mathcal E}_{\omega}|Lc/\omega
<< \phi_{0}.
\end{eqnarray}
We calculate up to the first order perturbation term of
 $a_{t-\eta/2}(\eta^{\prime}) =a(t-\eta/2-\eta^{\prime}/2)+a(t-\eta/2+\eta^{\prime}/2)$
in ${\bold F}_{t-\eta/2}$;
\begin{eqnarray}
\nonumber
&& {\bold F}_{t-\eta/2}(x,\eta;x,-\eta)
\\ \nonumber
&=&
{\bold F}^{(0)}_{t-\eta/2}(x,\eta;x,-\eta)
+\int dx^{\prime} \int_{-\eta}^{\eta}d\eta^{\prime}
 {\bold F}^{(0)}_{t-\eta/2}(x,\eta;x^{\prime},\eta^{\prime})
\\&\times&
\left( v_{F}\frac{ie}{\hbar c}\sigma_{z}
a_{t-\eta/2}(\eta^{\prime})\right)
 {\bold F}^{(0)}_{t-\eta/2}(x^{\prime},\eta^{\prime};x,-\eta)+\cdots,
\end{eqnarray}
 where ${\bold F}^{(0)}_{t-\eta/2}(x^{\prime},\eta^{\prime};x,-\eta)$
denotes the ${\bold F}$ matrix in the absence of time-dependent field, $a_{\omega} = 0$.
After a long but straightforward calculation, we get
the expression for the DC current:
\begin{eqnarray}
\nonumber
&&I_{0}=C_{\beta}\frac{|e|}{\tau_{D}}
\left( \frac{\phi_{\omega}}{\phi_{0}} \right)^{2}
\times
\\
\nonumber
&&\sum_{m=-\infty}^{\infty}
\frac{4\pi^{2} (\omega \tau_{D})^{2}k_{m}  }
{\left[
\big(k_{m}^{2}-(\omega \tau_{f})^{2}
+\tau_{D}/\tau_{\phi}^{*} \big)^{2}
+(\omega \tau_{D})^{2}\right]
\left[ k_{m}^{2}
+\tau_{D}/\tau_{\phi}^{*}\right]}
\\ &&\times
\left( 1-\frac{\tau}{\tau_{D}}
\big(k_{m}^{2}-(\omega \tau_{f})^{2}+
 \tau_{D}/\tau_{\phi}^{*} \big)
\right),
\label{i_weak}
\end{eqnarray}
where
$k_{m}=2\pi(m+2\phi/\phi_{0})$
($\phi=\bar{A}L$ is the static magnetic flux),
$\tau_{D}=L^{2}/D$ is the diffusion time, and
$\tau_{f}=L/v_{F}$ is a ballistic time scale.
By neglecting ballistic parameters in Eq.(\ref{i_weak})
 i.e,  $ \tau/\tau_{D}\rightarrow 0$, and
 $\omega \tau_{f} =\omega \tau_{D} \sqrt{\tau/\tau_{D}}\rightarrow 0$,
  we recover the
earlier result of $I_{0}$ by Kravtsov and Yudson\cite{kravtsov1}.
Compared with the results for diffusive limit\cite{kravtsov1}, we basically encounter a new parameters $\tau/\tau_{D}$ by
considering ballistic effects.

Since $I_{0}$ is periodic with a period
$\phi_{0}/2$, the Fourier components $I^{(n)}$of $I_{0}$ are often the quantities
under study;
\begin{eqnarray}
I_{0}(\phi)=C_{\beta}\frac{|e|}{\tau_{D}}\sum_{n} I^{(n)}\sin (4\pi n
\frac{\phi}{\phi_{0}}).
\end{eqnarray}
In Fig. 1, we plot
the amplitude of the first harmonic $I^{(1)}$of $I_{0}(\phi)$
using Eq.(\ref{i_weak}).
In contrast with the current for diffusive limit\cite{kravtsov1}
,i.e. $\tau/\tau_{D}=0$, $I^{(1)}$ for finite $\tau/\tau_{D}$ show
oscillation behaviour.
A new time scale $\tau_{f}=L/v_{f}$ appears associated with the oscillation period $\Delta
\omega=2\pi/\tau_{f}$.
Note that when we take into account ballistic effects, (i.e.,  $\tau/\tau_{D}\neq 0$),
 we can not neglect $\omega\tau_{f}(= \omega\tau_{D}\sqrt{\tau/\tau_{D}})$
in the denominator of Eq.(\ref{i_weak}),
which gives oscillating behaviour in Fig. 1.
Intuitively, this oscillation is due to the fact that the time period of periodic orbits along the ring
match with that of the applied external field.

Now, let us look into a different regime where the disorder
potential is very weak but the applied field is arbitrarily
strong;
\begin{eqnarray}
\frac{1}{\tau} \ll \omega,~~~{\rm and}~ \frac{1}{\tau_{f}}.
\end{eqnarray}
For this case, we use perturbation of the electron-impurity sacttering term $\sigma_{x}/\tau$ with
parameter $1/\omega\tau << 1$.
The leading terms are written as
\begin{eqnarray}
\nonumber
I^{(n)}&\approx &{\mathcal F}_{n}(\pi \frac{\phi_{\omega}}{\phi_{0}},\omega \tau_{f} ) e^{-n\frac{\tau_{f}}{\tau}}
\\ &\times& \left[ \sin(n\omega\tau_{f}/4) + \frac{1}{\omega\tau} \left( \frac{\cos(n\omega\tau_{f}/4)+(2/\omega\tau)
\sin(n\omega\tau_{f}/4) }{1+(2/\omega\tau)^{2}}  \right)+\cdots
\right],
\end{eqnarray}
where
\begin{eqnarray}
{\mathcal F}_{n}(x,y)=xy J_{1}(16x\sin(ny/4)/y).
\end{eqnarray}
Here $J_{1}$ is the Bessel function of order 1.

As shown in Fig. 2, the first harmonic $I^{(1)}$ of the current
may show sign reversal when the applied field is not too weak.
When the magnetic flux $\phi_{\omega}$ associated with the time
dependent field is larger than half flux quantum $\phi_{0}/2$, $I^{(1)}$
is in a regime of negative sign depending on the applied frequency.
Interestingly, this is also the condition that the applied field
can cause dephasing of electrons efficiently.

Experimental configuration in Ref.\cite{deblock} seems to be
promising for the observation of the ballistic effects we
have discussed here.
However, instead of metals, GaAs samples will be more promising to
show ballistic effects, where
the mean-free-path is usually order of $\mu m$.
Furthermore, both of the well-defined amplitude and frequency are
necessary for the comparision.
In case of the samples with GaAs, the applied field of frequency $\omega$
in order of THz may clearly show the ballistic effects we
discussed.

%For the soft potential with disorder where $\omega\tau \gg 1$,  it
%should be noted that the effects we discussed is
%non-universal\cite{altland} i.e., the results are valid in a
%particular geometry of ring.
%In reality, there might be additional effects due to
%electron heating, finite cross-sections of the wires, and
%the non-monochromaticity of the applied field, which we have not
%discussed in this work.
%When the applied field is not monochromatic but contains a wide
%spectrum of frequency, we may expect the applied field to 
%play a destructive role of noise
%suppressing the coherence of electrons.
%It is not easy to define the role of a strong noisy field
%on the induced current due to the associated non-lnear properties.
%There might be a significant signal of the induced current due to the
%so-called non-linear
%effect ``stochastic resonance"\cite{ahn}.

In conclusion, we have shown that
the mesoscopic electron transport in disordered quantum wires
is described by a
generalized Cooperon equation which coincides in form with
Dirac equation for massive fermions in 1+1 dimensional system.
Ballistic effects in a disordered wire are equivalent to
the relativistic effects in clean one-dimensional systems.
Based on the new Cooperon equation,
electric currents in mesoscopic rings
induced by oscillating magnetic fields are calculated.
It is predicted that, as a ballistic effect, the DC component of the induced
electric currents shows oscillating behavior in the domain of
external-field frequency.
Furthermore, in the high frequency regime, the sign of the induced current 
can be either diamagnetic or
paramagnetic depending on the strength and the frequency of the
field.

The author would like to thank C. Kim, P. Mohanty, L.I. Glazman, B.I. Halperin,
D. Kim, M. Das, 
F. Green, Y. D. Park, C. Lee, M. Y. Choi, and
T. Yamamoto for stimulating discussions.

\begin{figure}
\centerline{\psfig{figure=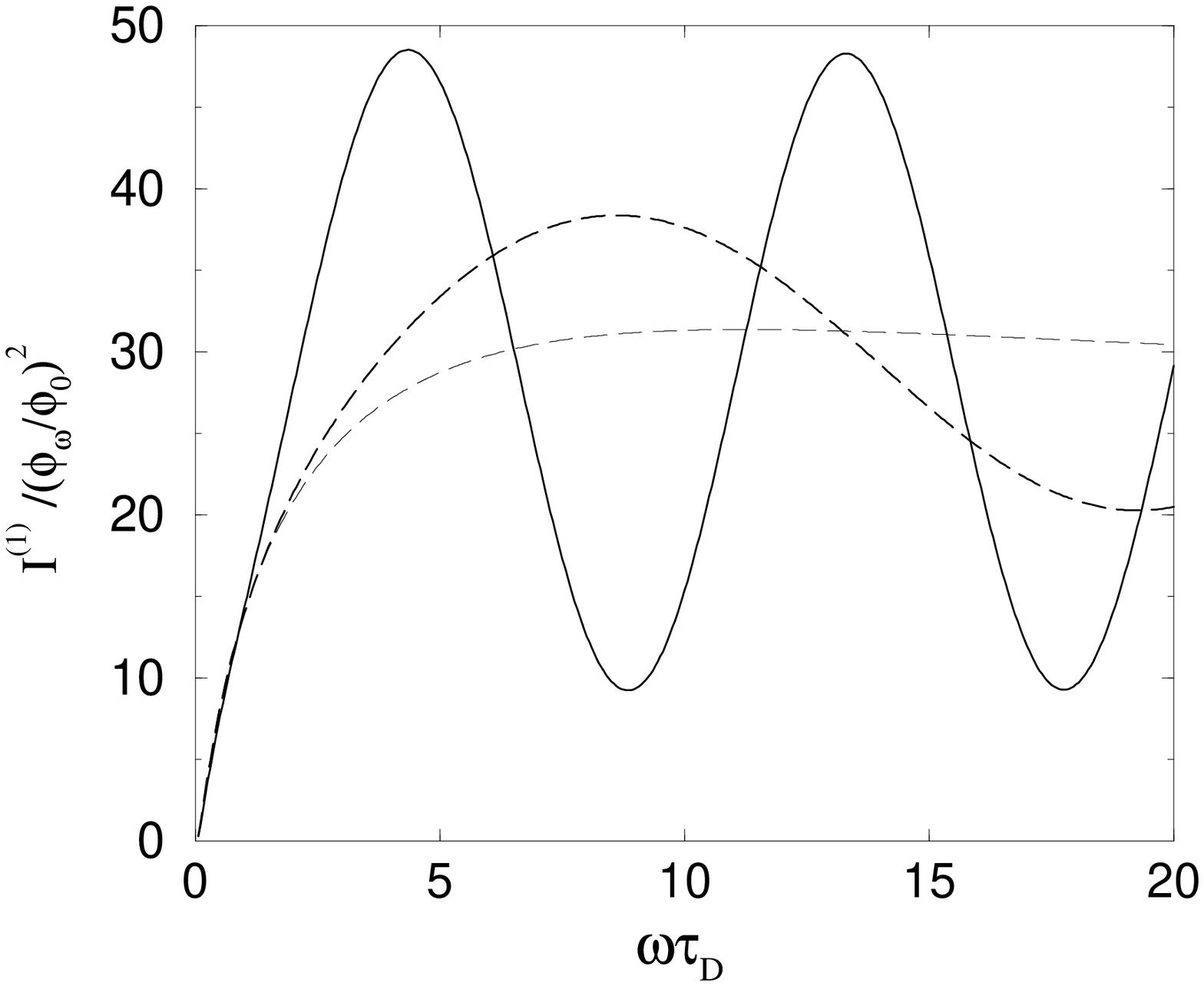,width=0.6\textwidth}}
 \caption{ The amplitude $I^{(1)}$ of the first harmonic
of $I_{0}(\phi)$ in units of $C_{\beta}|e|/\tau_{D}$ for weakly time-dependent field $\phi_{\omega} \ll \phi_{0}$
and different ``ballisticity" $\tau/\tau_{D}=0.5$ (thick solid line), $\tau/\tau_{D}=0.1$ (thick dashed line), and
$\tau/\tau_{D}=0$ (dashed line, Ref.[5]). $\tau_{\phi}^{*}$ was chosen to be $10 \tau_{D}$ for
all cases.  }
 \label{amplitude1}
\end{figure}

\begin{figure}
\centerline{\psfig{figure=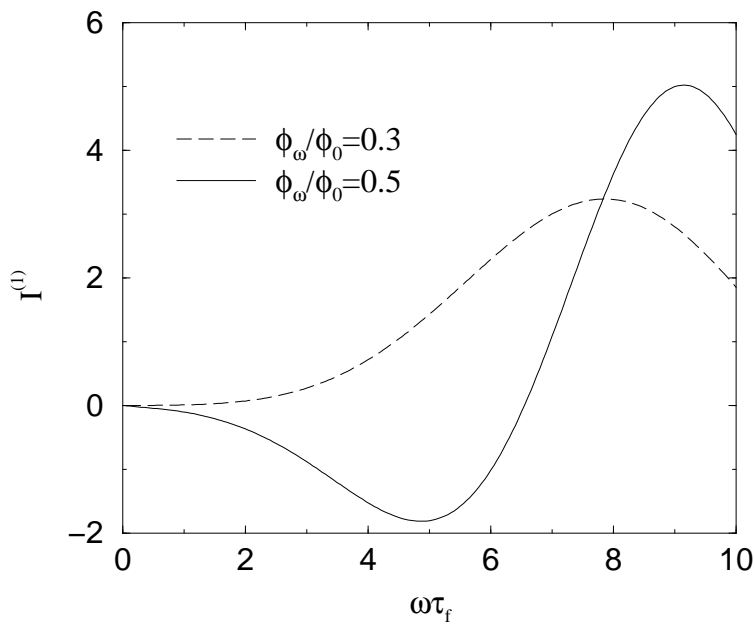,width=0.6\textwidth}}
 \caption{ The amplitude $I^{(1)}$ of the first harmonic
of $I_{0}(\phi)$ in units of $C_{\beta}|e|/\tau_{D}$ using high-frequency approximation
$ 1/\tau \ll \omega$,$1/\tau_{f}$.   $\tau_{f}/\tau=0.1$ was
chosen as a specific case.}
 \label{amplitude2}
\end{figure}

\end{document}